\documentclass[sigconf,screen]{acmart}
\usepackage[utf8]{inputenc}
\usepackage{multirow}
\usepackage{threeparttable}
\usepackage{makecell}
\usepackage{colortbl}
\usepackage{caption}
\usepackage{subcaption}
\usepackage{float}
\usepackage{xspace}
\usepackage{xcolor}
\usepackage{url}
\usepackage{enumitem}
\usepackage{wrapfig}
\usepackage{booktabs}
\usepackage{tabularx}
\usepackage{tcolorbox}
\tcbuselibrary{breakable}

\usepackage{amssymb}
\usepackage{hyperref}
\usepackage[linesnumbered,ruled,vlined]{algorithm2e}
\usepackage{amsmath}
\usepackage{wrapfig}
\usepackage{tabularx}
\usepackage{mathtools}
\usepackage{pbalance}

\AtBeginDocument{%
  }

\newcolumntype{C}[1]{>{\centering\arraybackslash}p{#1}}
\AtBeginDocument{%
  }

\setcopyright{cc}
\setcctype{by}
\acmDOI{10.1145/3832783.3837490}
\acmYear{2026}
\copyrightyear{2026}
\acmISBN{979-8-4007-2882-2/2026/10}
\acmConference[ASE '26]{Proceedings of the 41st IEEE/ACM International Conference on Automated Software Engineering}{October 12--16, 2026}{Munich, Germany}
\acmBooktitle{Proceedings of the 41st IEEE/ACM International Conference on Automated Software Engineering (ASE '26), October 12--16, 2026, Munich, Germany}
\acmSubmissionID{ase26main-p1325-p}
\received{2026-03-26}
\received[accepted]{2026-06-18}

\author{Ruijie Zhang}
\authornote{Ruijie Zhang, Huaxiao Yin, and Li Hu contributed equally to this research.}
\orcid{0009-0009-3379-4090}
\email{jeffy353866@gmail.com}
\affiliation{%
  \institution{Institute of Information Engineering, Chinese Academy of Sciences}
  \city{Beijing}
  \country{China}
}
\affiliation{%
  \institution{School of Cyber Security, University of Chinese Academy of Sciences}
  \city{Beijing}
  \country{China}
}

\author{Huaxiao Yin}
\authornotemark[1]
\orcid{0009-0001-7289-2525}
\email{beautysheep4021@163.com}
\affiliation{%
  \institution{Institute of Information Engineering, Chinese Academy of Sciences}
  \city{Beijing}
  \country{China}
}
\affiliation{%
  \institution{School of Cyber Security, University of Chinese Academy of Sciences}
  \city{Beijing}
  \country{China}
}

\author{Li Hu}
\authornotemark[1]
\authornote{Li Hu and Kun Zhang are the corresponding authors.}
\orcid{0009-0001-7857-7454}
\email{pdxbshx@mail.ustc.edu.cn}
\affiliation{%
  \institution{University of Science and Technology of China}
  \city{Hefei}
  \country{China}
}

\author{Fanqi Kong}
\orcid{0009-0002-7793-8112}
\email{17358190815@163.com}
\affiliation{%
  \institution{Peking University}
  \city{Beijing}
  \country{China}
}

\author{Haozhe Liang}
\orcid{0009-0007-4210-160X}
\email{lianghaozhe2024@163.com}
\affiliation{%
  \institution{University of Chinese Academy of Sciences}
  \city{Beijing}
  \country{China}
}

\author{Shaoyin Cheng}
\authornote{Also affiliated with Anhui Province Key Laboratory of Digital Security, Hefei, China.}
\orcid{0000-0002-3992-9509}
\email{sycheng@ustc.edu.cn}
\affiliation{%
  \institution{University of Science and Technology of China}
  \city{Hefei}
  \country{China}
}

\author{Kun Zhang}
\authornotemark[2]
\orcid{0000-0003-2278-0979}
\email{zhangkun@iie.ac.cn}
\affiliation{%
  \institution{Institute of Information Engineering, Chinese Academy of Sciences}
  \city{Beijing}
  \country{China}
}
\affiliation{%
  \institution{School of Cyber Security, University of Chinese Academy of Sciences}
  \city{Beijing}
  \country{China}
}

\author{Bibo Tu}
\orcid{0000-0002-0278-7420}
\email{tubibo@iie.ac.cn}
\affiliation{%
  \institution{Institute of Information Engineering, Chinese Academy of Sciences}
  \city{Beijing}
  \country{China}
}
\affiliation{%
  \institution{School of Cyber Security, University of Chinese Academy of Sciences}
  \city{Beijing}
  \country{China}
}

\newcommand{\sysbench}{{RegoBench}\xspace}
\newcommand{\sysmethod}{{RegoMender}\xspace}
\newtcolorbox{rqanswerbox}{
  breakable,
  colback=gray!10,
  colframe=gray!60,
  arc=1mm,
  boxrule=0.5pt,
  left=4pt,
  right=4pt,
  top=4pt,
  bottom=4pt,
  width=\linewidth,
  before skip=0.5ex,
  after skip=0.5ex
}

\begin{document}

\title{RegoMender: A Multi-agent Framework for Automated Rego Policy Repair}

\begin{abstract}
Policy-as-Code (PaC) has emerged as a foundational paradigm in cloud-native systems. However, repairing Rego policies remains challenging due to their strict syntax, declarative semantics, and specific verification requirements. Existing automated program repair (APR) methods are mainly designed for mainstream languages and struggle to accommodate the unique policy logic and repair workflows of Rego. Thus, Rego policy repair remains a manual process driven by trial and error. To address this, we introduce \sysbench, the first automated Rego policy benchmark comprising 1,159 real-world defects, and propose \sysmethod, a feedback-guided multi-agent framework. Collaborating via a shared blackboard and an atomic toolkit, \sysmethod dynamically executes patching, rewriting, or rolling back, to resolve complex repair scenarios. Evaluated on \sysbench, \sysmethod achieves syntactic and test-based semantic success rates of 82.9\% and 77.9\%, respectively, outperforming the strongest baseline by 13.4\%. Unlike other approaches whose performance drops in complex scenarios, \sysmethod maintains effectiveness across single-step and chained repair tasks. Furthermore, it generalizes strongly across HCL and SQL and achieves an 86.4\% repair rate on 132 real-world CVE vulnerabilities, proving its robustness and practical utility for security.
\end{abstract}

\vspace{-2.0ex}




\maketitle

\vspace{-2.0ex}
\section{Introduction}

Policy-as-Code has become a core paradigm for enforcing security and compliance requirements in cloud-native systems. It represents policies as executable rules that are continuously evaluated against structured artifacts, such as API requests, configuration objects, and infrastructure-as-code (IaC) plans, across both delivery and runtime environments. A widely adopted implementation of PaC is Open Policy Agent (OPA), which separates policy logic from application code and provides a unified evaluation framework across heterogeneous platforms. In practice, OPA is integrated into CI/CD pipelines for pre-deployment validation and into Kubernetes admission control to prevent non-compliant resource updates. OPA policies are written in Rego, a declarative language designed to express policy rules over inputs and produce authorization decisions~\cite{opa_policy_language}.

However, repairing Rego policies is particularly challenging because of their declarative semantics and strict validation requirements. Unlike imperative languages, Rego follows a logic programming paradigm derived from Datalog, where unification combines pattern matching with variable binding. 
This context sensitivity means that even a seemingly simple modification may unintentionally alter the logic of the policy. Furthermore, policy evaluation may return \texttt{undefined} when certain data paths are missing. Rather than raising an error, this result may cause conditions to be skipped.

Meanwhile, the current APR benchmarks~\cite{ye2025llmsecconfig,saavedra2025infrafix,just2014defects4j,jimenez2024swebench} have been developed primarily for mainstream programming languages and rely on test suites as correctness oracles. Classic datasets such as Defects4J~\cite{just2014defects4j} provide standardized collections of real-world Java bugs with regression tests, while repository-level benchmarks like SWE-bench~\cite{jimenez2024swebench} evaluate issue resolution in large-scale Python repositories, collectively advancing research in APR. However, dedicated benchmarks for evaluating the automated repair of Rego policies are absent. Unlike conventional programs, policies written in Rego are validated through a tool-driven workflow that includes linting and compilation, policy-specific testing, and compliance checks derived from domain requirements.

This mismatch presents two fundamental challenges for the automated repair of Rego policies: \textbf{(i) Lack of benchmarks for Rego repair.} To date, there is no publicly available dataset specifically designed for Rego policy repair, making it difficult to systematically evaluate whether existing APR or LLM-based approaches can effectively handle Rego-specific syntax, semantics, and verification workflows. \textbf{(ii) Heavy reliance on manual or ad hoc LLM-based repair workflows.} In practice, Rego policy repair still relies heavily on manual debugging by domain experts or on ad hoc interactions with LLMs
, developers must manually interpret policy failures, construct prompts, invoke \texttt{opa check} and \texttt{opa test}, assess the verification outcomes, and iteratively revise candidate patches. This human-in-the-loop repair process is time-consuming, error-prone, and difficult to scale, especially when repairs must preserve repository-specific policy intent and compliance semantics.

To address the lack of systematic data and reproducible evaluation for Rego policy repair, we mine real-world defect fixes from 2,847 GitHub repositories that actively maintain Rego policies. We then design a strict three-stage filtering pipeline to progressively refine the raw commit-level candidates, including isolated patch extraction, redundancy and noise mitigation, and deterministic verification under a standardized OPA toolchain. This process yields 1,159 high-quality buggy-fix pairs that are fully reproducible. This benchmark covers seven OPA-derived defect types, preserves the natural difficulty distribution of real-world defects and features a high proportion of complex chained repairs, directly measuring how well repair methods handle practical programming tasks.

Furthermore, we design \sysmethod, a multi-agent framework for automated Rego policy repair. \sysmethod establishes a closed loop between code repair and OPA verification by coordinating three specialized roles, namely the Orchestrator, Analyzer, and Repairer, through a shared Blackboard and an auditable atomic toolkit. Specifically, all repository interactions are carried out through the atomic toolkit, which provides five core operations: \texttt{Verify}, \texttt{Context Scan}, \texttt{Error Triage}, \texttt{Rewrite}, and \texttt{Patch}. This design ensures that each repair step remains precise and reproducible. Guided by real-time verifier feedback, the Orchestrator coordinates the Analyzer and Repairer, records progress on the Blackboard, and rolls back when necessary to avoid stagnation. In this way, \sysmethod supports reliable multi-step repair, particularly in cases where resolving one failure exposes additional hidden defects.

To evaluate the effectiveness of \sysmethod, we conduct extensive experiments on \sysbench. First, the results show that, compared to existing automated program repair methods~\cite{yin2024thinkrepair,bouzenia2025repairagent,xia2024agentless,hu2025aprmcts}, \sysmethod achieves syntactic and test-based semantic success rates of 82.9\% and 77.9\%, respectively. It 
ranks first across all seven defect types, demonstrating the effectiveness of our agent orchestration and task decision-making mechanisms. Second, we analyze the impact of different backbone models on \sysmethod and the baselines. We find that \sysmethod demonstrates superior robustness across different models
, and DeepSeek-V3.1-Terminus~\cite{liu2024deepseek} provides the best balance between performance and cost. Third, we conduct a comparative study on single-step and chained repair tasks. We find that 
\sysmethod maintains stable performance in both situations. Furthermore, we conduct an ablation study on the designed atomic toolkit. The results reveal that the \texttt{Context Scan}
tool has the most significant impact on the framework's performance. This indicates that the repair process for the Rego policies relies on context gathering
. To evaluate its cross-language generalizability, we extend our method to two additional domain-specific languages (HCL and SQL). The results show that \sysmethod continues to maintain a significant advantage over other baselines. Finally, to verify its practical utility in the real world, we evaluate our method on 132 CVE vulnerabilities. 
RegoMender achieves a repair rate of 86.4\%, outperforming ThinkRepair and RepairAgent and demonstrating its practical potential for security-related repair.

\begin{figure*}[t]
  \centering
  \captionsetup{skip=4pt}
  \scalebox{0.95}{
    \includegraphics[width=0.90\textwidth]{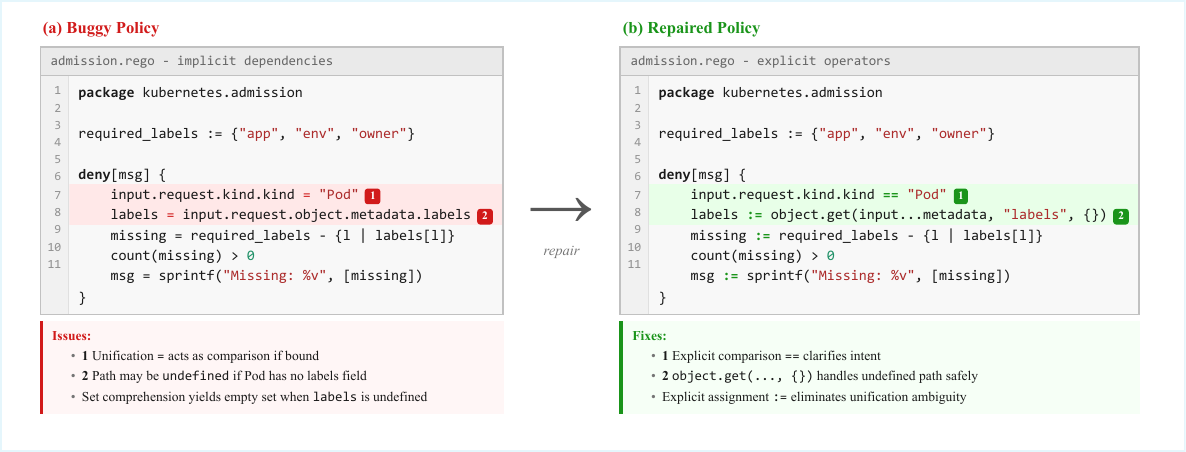} 
  }
  \vspace{-3.0ex}
  \caption{An Example for Rego Policy Repair.}
  \label{fig:intro}
  \vspace{-4.0ex}
\end{figure*}

In summary, our contributions are as follows:
\begin{itemize}[leftmargin=2em, nosep, topsep=0pt, parsep=0pt, itemsep=0pt]
  \item To the best of our knowledge, this is the first work to explore automated repair of Rego policies, laying the groundwork for systematic and reproducible research in this area.
  \item We introduce \sysbench, the first benchmark for automated Rego policy repair. To construct it, we design a three-stage filtering pipeline and apply it to 2,847 GitHub repositories that actively maintain Rego policies, curating a high-quality dataset of 1,159 real-world defects.
  \item We propose \sysmethod, a multi-agent framework where Orchestrator, Analyzer, and Repairer agents collaborate via a Blackboard and an auditable toolkit to achieve Rego repairs with rollback-based stagnation control.
  \item Extensive evaluations demonstrate that \sysmethod achieves state-of-the-art test-based semantic success rates of 77.9\%, outperforming 
  the strongest baseline by 13.4\%. Notably, it exhibits exceptional stability in complex multi-step defects and successfully resolves 86.4\% of real-world CVEs.
\end{itemize}

\vspace{-2.0ex}
\section{Motivation}
\label{motivation}
\vspace{-0.3ex}

Figure~\ref{fig:intro} shows a motivating example of Rego policy repair. In Figure~\ref{fig:intro}(a), the buggy rule uses unification (\texttt{=}) and directly reads \texttt{input.request.object.metadata.labels} after checking that the resource kind is \texttt{"Pod"}. Under Rego's undefined-value semantics, if the labels field is absent or the binding fails, the rule body may become \texttt{undefined}, so the intended deny decision is not produced. In Figure~\ref{fig:intro}(b), the repair uses explicit comparison (\texttt{==}) for the kind check, retrieves labels with \texttt{object.get(input...metadata, "labels", {})}, and uses strict assignment (\texttt{:=}) for derived values. These changes handle missing labels explicitly and restore a well-defined admission decision.
This example shows why Rego repair is language-specific. The operators \texttt{=}, \texttt{:=}, and \texttt{==} represent unification, assignment, and comparison rather than simple symbol replacements, and whether an input path is defined depends on the runtime context. Therefore, Rego repairs should be validated with the standard OPA toolchain (\texttt{opa check} and \texttt{opa test}) to avoid deployment failures. Current LLMs also have limited experience with Rego repair because repair data remains scarce, making reliable decisions from prior patterns difficult.

Based on the above analysis, we construct \sysbench, a benchmark of 1,159 Rego repair instances covering a broad range of syntax and semantic errors, and propose \sysmethod, a multi-agent framework for automated Rego policy repair. Recent software engineering systems use specialized agents for planning, analysis, coding, and tool use~\cite{he2024mas4se_survey,qian2024chatdev,hong2023metagpt,tufano2024autodev}. However, such collaboration does not directly address Rego-specific needs in policy semantics, context extraction, and OPA-based verification. To address these needs, our framework coordinates three specialized agents via a shared toolkit and blackboard for automated Rego policy repair.

\section{Overview}

In this section, we first describe the construction of the proposed benchmark, \sysbench (\S\ref{regobench}), followed by a detailed presentation of our multi-agent repair framework, \sysmethod (\S\ref{regomender}).

\subsection{\sysbench Construction}
\label{regobench}

We construct \sysbench, the first benchmark dedicated to real-world Rego defects, follows three main steps. First, we define each repair instance and the evaluation criteria for a successful repair. Next, we execute the data filtering process to collect high-quality instances from open-source repositories. Finally, we classify the benchmark along several characteristic dimensions to provide a comprehensive overview of the dataset.

\begin{figure*}[tbp]
  \centering
  \captionsetup{skip=4pt}
  \scalebox{0.85}{
    \includegraphics[width=0.95\textwidth]{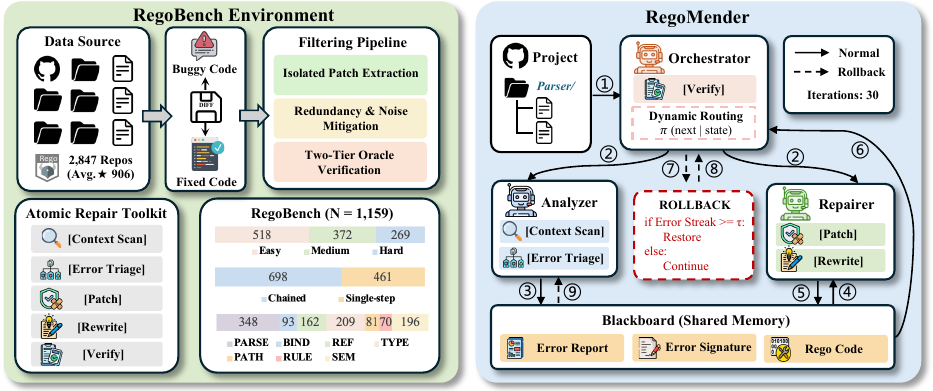}
  }
  \vspace{-0.5ex}
  \caption{
  Overview of RegoBench Construction and RegoMender Workflow.
  }
  \label{fig:main}
  \vspace{-3.5ex}
\end{figure*}

\subsubsection{\textbf{Instance Definition.}}
\label{sec:instance}

We define each repair instance through four core components that encapsulate all necessary context, ensuring each defect is self-contained and independently verifiable:

\begin{itemize}[leftmargin=2em, nosep, topsep=0pt, parsep=0pt, itemsep=0pt]
  \item \textbf{Code Pair.} We store two versions of the policy code: $C_{bug}$ and $C_{fix}$. $C_{bug}$ represents the defective policy, while $C_{fix}$ is the corresponding manual fix. Alongside the raw code, we record metadata, such as the commit hash, repository link, and relative file paths. This information enables the precise historical code state to be retrieved and replayed deterministically.
  \item \textbf{Controlled Context.} This component includes dependency resolution results, the module import graph, and generated code stubs. These elements allow for isolated verification, meaning the policy can be evaluated locally without the need to download or configure the entire repository.
  \item \textbf{Standardized Toolchain.} We enforce a fixed OPA version (v1.10.1) and consistent command line arguments. This ensures consistent diagnostic messages and removes any variability that might arise from using different tool versions.
  \item \textbf{Two-Tier Oracle Records.} We organize verification into two tiers. The first ensures syntactic correctness via \texttt{opa check}, covering syntax errors, linking issues, type mismatches, and safety violations. The second ensures semantic correctness via \texttt{opa test} to prevent logical regressions. To qualify as a valid repair instance, the code must pass both tiers, guaranteeing that every sample is syntactically and behaviorally correct.
\end{itemize}

\subsubsection{\textbf{Data Filtering.}}

To construct the initial corpus, we query the GitHub Search API for repositories containing \texttt{.rego} files. To ensure data quality and relevance, we apply a strict activity filter, retaining only repositories with at least 40 commits between January 1, 2020, and June 30, 2024. This yields 2,847 qualified repositories. We then analyze their commit histories to extract potential defect repairs, resulting in 54,821 raw candidates. 
However, extracting these defects from open source repositories introduces noise. Specifically, we observe four patterns of invalid repairs. First, superficial fixes occur when commits labeled as repairs consist only of formatting changes or code refactoring. Second, invalid repairs occur when developers simply delete code to pass checks without fixing the actual defect. Third, coupling across files scatters the repair logic across multiple formats, such as Rego, YAML, and Go, making it difficult to isolate the core defect. Fourth, environmental instability arises when failures depend on external variables or continuous integration contexts that are difficult to reproduce. 
We therefore design a three-stage filtering pipeline to obtain a high-quality dataset.

\textit{1) \underline{Isolated Patch Extraction.}} To resolve coupling across files, we enforce a strict single-file pairing rule. Only commits with substantive edits confined to a single \texttt{.rego} file are retained, ensuring a clear mapping between the defective and fixed states at the same file path. Nonfunctional edits
are discarded to ensure each instance represents a meaningful repair. This filtering step reduces the raw corpus from 54,821 to 8,643 candidates.

\textit{2) \underline{Redundancy \& Noise Mitigation.}} To reduce noise from repetitive patches, we implement a clear deduplication strategy. We first exclude forked and archived repositories and then use file fingerprints, defined by the repository, commit hash, and file path, to prevent redundant processing. Additionally, we apply content hashing to eliminate identical code copies caused by version control operations. Finally, we manually audit the dataset to discard samples where the link between the code change and the verification result is unclear. Through this process, we retain 3,071 candidates.

\textit{3) \underline{Two-Tier Oracle Verification.}} Finally, to ensure reproducibility and eliminate environmental inconsistencies, we establish a deterministic two-tier verification process using the standard OPA toolchain. Each retained instance must satisfy strict evaluation criteria. First, the defective policy must trigger a verifiable failure (via either \texttt{opa check} or \texttt{opa test}) to confirm the presence of a bug. Second, the developer fix must successfully pass both verification tiers. This process guarantees that every accepted instance is both syntactically valid and behaviorally correct. 

Following this three-stage filtering pipeline, we refine the raw data into a dataset of 1,159 high-quality and reproducible instances.

\begin{table}[t]
\centering
\captionsetup{
    skip=4pt
}
\renewcommand{\arraystretch}{0.90}
\setlength{\tabcolsep}{2.2mm}

\caption{Defect Taxonomy of RegoBench.}
\label{tab:defect_taxonomy_simple}
\vspace{-0.5ex}

\scalebox{0.83}{
    \begin{tabular}{llc}
    \toprule
    \textbf{Defect Type} & \textbf{Symptom} & \textbf{Number} \\
    \midrule
    PARSE & parse error / rule-head / Rego.v1 & 348 \\
    BIND  & binding operator / scope & 93 \\
    REF   & unsafe var / undefined ref / import mismatch & 162 \\
    TYPE  & type error / arity / built-in / helper contract & 209 \\
    PATH  & missing path / input shape & 81 \\
    RULE  & recursion / conflicting rules / defaults & 70 \\
    SEM   & \texttt{opa test} failure after \texttt{opa check} & 196 \\
    \midrule
    \textbf{Total} & --- & \textbf{1,159} \\
    \bottomrule
    \end{tabular}
}  
\vspace{-4.0ex}
\end{table}

\vspace{-1.5ex}
\subsubsection{\textbf{Dataset Characterization.}}
\label{sec:category}

Following standard practices~\cite{jimenez2024swebench,just2014defects4j,le2015manybugs,saha2018bugs} in benchmark construction, we preserve the natural distribution of the filtered dataset. To characterize this dataset, we classify the repair difficulty into three levels: easy, medium, and hard. We then engage three senior developers to manually evaluate and classify the 1,159 repair instances. To measure the consistency of their independent evaluations, we calculate the Fleiss' Kappa~\cite{fleiss1971measuring} statistic. The resulting score is 0.85, which indicates an almost perfect agreement among the three developers.
For the few instances with conflicting classifications, the developers hold discussions to review the instances and reach a final consensus.
The labeling results reveal that 518 instances fall into the easy category, 372 into the medium category, and 269 into the hard category. This proportion is consistent with previous empirical findings~\cite{karampatsis2020often,purushothaman2005toward,zhong2015empirical}, confirming that real software defects naturally concentrate on simpler errors. 

Beyond repair difficulty, we characterize the benchmark by repair dependency, categorizing instances as single-step or chained repairs. Specifically, an instance is classified as single-step if the initial error category remains unchanged until the repair successfully passes both evaluation tiers. In contrast, an instance is labeled as chained if resolving the initial error exposes at least one subsequent error category before overall success. Based on this error-driven definition, 698 instances are chained repairs and the remaining 461 are single-step repairs, as shown in Figure~\ref{fig:main}.

We further characterize each RegoBench instance by its initial defect type. Based on OPA verification feedback, we define seven categories: PARSE, BIND, REF, TYPE, PATH, RULE, and SEM. For a chained repair, the defect type is determined by the first error exposed during verification. Table~\ref{tab:defect_taxonomy_simple} presents the definition, typical symptoms, and number of instances for each category.

\subsection{RegoMender Framework}
\label{regomender}
\vspace{-0.3ex}

RegoMender is an automated multi-agent framework designed for repairing Rego policies. As shown in Figure~\ref{fig:main}, our approach centers on an Orchestrator that manages a continuous feedback loop driven by the results of \texttt{opa check} and \texttt{opa test}. The Orchestrator coordinates two specialized agents, namely the Analyzer and the Repairer, using a shared blackboard to exchange state information. To make precise code changes and prevent the repair process from getting stuck, RegoMender includes an atomic repair toolkit, as well as a mechanism for stagnation detection and rollback. We detail these core components and the overall execution workflow below.

\vspace{-1.0ex}
\subsubsection{\textbf{Components.}} Each part RegoMender takes on a distinct role to ensure a smooth repair process. We outline their specific tasks and internal logic below.

\textit{\underline{Atomic Repair Toolkit.}} To make the system behavior auditable and reproducible, we abstract all operations into five atomic tools. These tools serve as the sole interface for the agents:
\begin{itemize}[leftmargin=2em, nosep, topsep=0pt, parsep=0pt, itemsep=0pt]
    \item \textbf{Verify:} Provides the system source of truth by running syntax and semantic checks to return a formal set of errors.
    \item \textbf{Context Scan:} Extracts an abstract syntax tree and a dependency graph from the current policy to help agents understand the available variables and packages.
    \item \textbf{Error Triage:} Converts raw error information into a structured format containing the error category, line location, and the specific message associated with each line.
    \item \textbf{Patch:} Conducts corrections on individual code lines, such as modifying a variable name, flipping a logical operator, or refining a specific policy constraint.
    \item \textbf{Rewrite:} Replaces entire policy blocks or regenerates complex logic when simple edits are insufficient.
\end{itemize}

\textit{\underline{Blackboard.}} To reduce context loss in multi-agent collaboration, \sysmethod uses a structured shared blackboard as the central repository for the system state. This component contains three fields: (i) the Code Field, which maintains a versioned history of the Rego code, allowing the system to retrieve previous code snapshots for rollback; (ii) the Report Field, which stores the sequence of structured diagnostic reports generated by the Analyzer, including the error category, line location, corresponding error message, relevant context information extracted by the \texttt{Context Scan} tool, and a priority value to guide the Repairer in fixing critical issues first; and (iii) the Error Signature Field, which maintains a tuple $\langle \textit{Error Category}, \textit{Error Count}, \textit{Error Streak} \rangle$. We define the first two elements as a state $(state = \langle \textit{Error Category}, \textit{Error Count} \rangle)$, which serves as the main signal for agent routing. The \textit{Error Streak} then records the number of consecutive occurrences of the same $state$. This allows the Orchestrator to monitor the persistence of failure patterns, enabling it to 
trigger rollbacks or strategy adjustments.

\textit{\underline{Orchestrator.}}
The Orchestrator serves as the decision-maker and quality controller of RegoMender. It does not generate code directly but coordinates the workflow by dispatching tasks to the Analyzer or the Repairer. Its strategy is determined by a decision rule $\pi(\mathit{next} \mid \mathit{state})$, where $\mathit{next} \in \{\mathit{Analyzer}, \mathit{Repairer}, \mathit{Rollback}, \mathit{Done}\}$ and state captures the current error category and count.
In each iteration, the Orchestrator retrieves the latest code from the Blackboard and invokes the \texttt{Verify} tool. 
If verification fails, the Orchestrator updates the Error Signature and selects the next action based on the current repair state and explicit routing conditions defined in its prompt. Localized failures with clear line-level diagnostics are treated as simple errors and sent directly to the Repairer. Ambiguous, dependency-related, or repeatedly unresolved failures are treated as complex errors and sent to the Analyzer for additional context and dependency analysis before repair. After each repair, the Orchestrator invokes Verify again. A changed state indicates progress and resets the Error Streak, whereas an unchanged state triggers a revised diagnosis from the Analyzer. Once the predefined stagnation threshold is reached, the Orchestrator rolls back to the snapshot saved before the stalled attempts and requests a new diagnosis.
The workflow concludes as $\mathit{Done}$ only after the policy passes both syntactic and semantic validation. To prevent infinite loops, the Orchestrator monitors the repair progress and stops the process if the maximum iteration limit is reached.

\textit{\underline{Analyzer.}} The Analyzer acts as the diagnostic specialist, engaged by the Orchestrator when error messages are too complex or ambiguous for direct repair. Upon receiving this task, it retrieves the latest error stream and uses the \texttt{Error Triage} tool to extract a structured error description, mapping each error to its category, line location, and corresponding message. At the same time, the \texttt{Context Scan} tool extracts context and dependency information for the identified code region. The Analyzer then analyzes each error to infer its underlying cause and synthesizes these findings into a diagnostic report written to the shared blackboard.

\textit{\underline{Repairer.}} As the execution specialist, the Repairer modifies the Rego code based on the diagnostic report retrieved from the Blackboard. 
It uses \texttt{Rewrite} tool for normalization or a high volume of structural issues (i.e., more than 10 errors), and \texttt{Patch} tool for localized faults. 
The resulting patch list specifies the affected lines and edits, after which the repaired code is synchronized back to the Blackboard for the next verification round.

\textit{\underline{Stagnation Detection and Rollback Mechanism.}} To prevent infinite loops caused by complex dependency chains, the Orchestrator continuously monitors repair progress. In our framework, stagnation is defined as a state where no meaningful progress is made. 
Specifically, we consider the repair progress stalled if the Error Signature remains unchanged for 5 consecutive iterations ($\tau = 5$). When this happens, the Orchestrator triggers a rollback by retrieving the Rego code snapshot from the Blackboard that existed before the stagnation began. It then dispatches the Analyzer to review the historical reports stored on the Blackboard and reflect on why the previous attempts failed. Instead of a standard analysis, the Analyzer re-evaluates the persistent errors and their dependencies to issue a revised diagnostic report. This allows the Repairer to change its approach and try a different repair path to find a successful solution.


\subsubsection{\textbf{Workflow.}} The Orchestrator manages \sysmethod as a feedback loop for repairing a faulty Rego repository. It extracts the initial code, stores it in the Code Field of the Blackboard, and invokes the \texttt{Verify} tool on the latest code in each iteration for syntactic and semantic testing. If all tests pass, the workflow terminates as \textit{Done}. Otherwise, the Orchestrator updates the Error Signature Field and selects the next action based on the current $state$. For simple errors, it dispatches the Repairer to use the \texttt{Patch} or \texttt{Rewrite} tool; for complex errors, it invokes the Analyzer to synthesize raw logs and dependency context into the Report Field. To prevent infinite loops, the Orchestrator tracks the \textit{Error Streak}. If the $state$ remains unchanged for 5 iterations, it rolls back to the snapshot saved prior to the stagnation. The Analyzer then reviews prior reports, revises the diagnostic strategy, and provides fresh guidance to the Repairer. This diagnosis, repair, and verification cycle continues until the policy is repaired or the process reaches the maximum limit of 30 iterations. We set the iteration limit and stagnation threshold based on our practical repair experience to balance repair effectiveness with computational efficiency.

\section{Experimental Setup}

We design a series of experiments to evaluate the effectiveness and efficiency of \sysmethod. Our study is guided by the following \textbf{Research Questions (RQs):}

\begin{itemize}[leftmargin=1em, nosep, topsep=0pt, parsep=0pt, itemsep=1pt]
  \item \textbf{RQ1:} How effective is \sysmethod in repairing Rego policies?
  \item \textbf{RQ2:} How do different LLM backbones impact the performance of \sysmethod?
  \item \textbf{RQ3:} How does \sysmethod utilize collaborative tool orchestration to resolve chained repairs?
  \item \textbf{RQ4:} How does each atomic tool contribute to \sysmethod?
  \item \textbf{RQ5:} To what extent does \sysmethod generalize to domain-specific language similar to Rego?
  \item \textbf{RQ6:} Can \sysmethod repair vulnerabilities in real-world deployment scenarios?
\end{itemize}

\subsection{Baselines}

Since automated Rego policy repair remains unexplored, we modify several existing automated repair methods~\cite{xia2024agentless,yin2024thinkrepair,hu2025aprmcts,bouzenia2025repairagent,hong2023metagpt,tufano2024autodev} to make them applicable to our study. We categorize them into standalone repair methods and general-purpose agent systems.

\textbf{Standalone Repair Methods.} This category includes state-of-the-art (SOTA) automated program repair work tailored for LLMs. Specifically, Agentless~\cite{xia2024agentless} is a lightweight approach that avoids complex agent loops by using a strict two-phase pipeline of fault localization and patch generation. ThinkRepair~\cite{yin2024thinkrepair} enhances patch quality by prompting the model to generate a structured thought process before outputting code. APR-MCTS~\cite{hu2025aprmcts} integrates LLMs with Monte Carlo Tree Search (MCTS)~\cite{browne2012survey} to explore possible patches using test feedback as reward signals, while RepairAgent~\cite{bouzenia2025repairagent} formulates bug fixing as an interactive task where the LLM autonomously gathers context to construct a repair.

\textbf{General-Purpose Agent Systems.} This category covers general-purpose agent systems designed for repository-level software engineering rather than specifically for automated program repair. It covers both multi-agent strategies and off-the-shelf coding agents. 
AutoDev~\cite{tufano2024autodev} uses a conversation manager to coordinate repository-level tasks and execute commands, while MetaGPT~\cite{hong2023metagpt} assigns specialized roles to agents to support collaboration similar to that of a software development team.
Claude Code~\cite{anthropic_claudecode2026} and Codex~\cite{openai_codex2026} are off-the-shelf coding agents that can inspect repository context, modify source files, and execute development tools. We evaluate both coding agents through their official SDKs.

We evaluate all baselines in RQ1. To manage costs, RQ2–RQ6 focus on two representative baselines: ThinkRepair and RepairAgent. To mitigate LLM randomness, we repeat each experiment three times and report the average performance.

\begin{table*}[t]
\centering
\captionsetup{skip=4pt}
\renewcommand{\arraystretch}{0.95}
\caption{Performance Comparison of Repair Methods on RegoBench.}
\setlength{\tabcolsep}{0.75mm}
\scalebox{0.85}{
\begin{threeparttable}
  \begin{tabular}{l|cccc|cccc|cccc|cccc}
  \toprule
      \multirow{2}{*}{\textbf{Method}} & \multicolumn{4}{c}{\textbf{Pass@Tier-1 (\%)}} & \multicolumn{4}{c}{\textbf{Pass@Tier-2 (\%)}} & \multicolumn{4}{c}{\textbf{JudgePass (\%)}} & \multicolumn{4}{c}{\textbf{Cost / Effort}} \\
      \cmidrule(lr){2-5} \cmidrule(lr){6-9} \cmidrule(lr){10-13} \cmidrule(lr){14-17}
      & \textit{Easy} & \textit{Medium} & \textit{Hard} & \textit{Overall} & \textit{Easy} & \textit{Medium} & \textit{Hard} & \textit{Overall} & \textit{Easy} & \textit{Medium} & \textit{Hard} & \textit{Overall} & \textit{Budget (\$)} & \textit{Time (h)} & \textit{LLM
Calls} & \textit{Regr.\%} \\
  \midrule
      \multicolumn{17}{l}{\textit{\textbf{Standalone Methods}}} \\
      Agentless~\cite{xia2024agentless} & 66.7{\tiny$\pm$1.8} & 60.0{\tiny$\pm$1.6} & 46.7{\tiny$\pm$1.9} & 59.9{\tiny$\pm$1.5} & 61.4{\tiny$\pm$1.7} & 54.0{\tiny$\pm$1.5} & 41.1{\tiny$\pm$1.8} & 54.3{\tiny$\pm$1.4} & 58.0{\tiny$\pm$1.6} & 50.2{\tiny$\pm$1.5} & 37.0{\tiny$\pm$1.6} & 50.6{\tiny$\pm$1.3} & 4.8{\tiny$\pm$0.2} & 2.6{\tiny$\pm$0.2} & 2,614{\tiny$\pm$74} & 17.5{\tiny$\pm$1.6} \\
        ThinkRepair~\cite{yin2024thinkrepair} & 71.1{\tiny$\pm$2.0} & 46.9{\tiny$\pm$1.8} & 27.8{\tiny$\pm$1.6} & 53.3{\tiny$\pm$1.4} & 65.1{\tiny$\pm$1.9} & 42.0{\tiny$\pm$1.6} & 24.1{\tiny$\pm$1.4} & 48.2{\tiny$\pm$1.1} & 60.2{\tiny$\pm$1.9} & 37.9{\tiny$\pm$1.5} & 20.9{\tiny$\pm$1.5} & 43.9{\tiny$\pm$1.2} & 3.9{\tiny$\pm$0.2} & 2.2{\tiny$\pm$0.1} & 1,998{\tiny$\pm$61} & 26.3{\tiny$\pm$1.8} \\
       APR-MCTS~\cite{hu2025aprmcts} & 73.3{\tiny$\pm$1.9} & 54.5{\tiny$\pm$1.7} & 46.7{\tiny$\pm$1.8} & 61.1{\tiny$\pm$1.5} & 67.4{\tiny$\pm$1.8} & 49.1{\tiny$\pm$1.6} & 41.1{\tiny$\pm$1.7} & 55.4{\tiny$\pm$1.7} & 64.0{\tiny$\pm$1.7} & 45.9{\tiny$\pm$1.5} & 37.0{\tiny$\pm$1.7} & 51.9{\tiny$\pm$1.6} & 6.2{\tiny$\pm$0.3} & 2.7{\tiny$\pm$0.2} & 4,995{\tiny$\pm$118} & 15.8{\tiny$\pm$1.4} \\
       RepairAgent~\cite{bouzenia2025repairagent} & 84.6{\tiny$\pm$1.6} & 64.2{\tiny$\pm$1.4} & 51.5{\tiny$\pm$1.5} & 70.4{\tiny$\pm$1.3} & 78.5{\tiny$\pm$1.5} & 58.4{\tiny$\pm$1.3} & 45.8{\tiny$\pm$1.4} & 64.5{\tiny$\pm$1.2} & 75.9{\tiny$\pm$1.4} & 55.4{\tiny$\pm$1.5} & 42.4{\tiny$\pm$1.4} & 61.5{\tiny$\pm$1.3} & 5.2{\tiny$\pm$0.2} & 2.9{\tiny$\pm$0.2} & 2,664{\tiny$\pm$72} & 14.8{\tiny$\pm$1.2} \\
  \midrule
      \multicolumn{17}{l}{\textit{\textbf{General-Purpose Agent Systems}}} \\
       MetaGPT~\cite{hong2023metagpt} & 72.5{\tiny$\pm$2.1} & 54.7{\tiny$\pm$1.8} & 41.1{\tiny$\pm$1.9} & 59.5{\tiny$\pm$1.6} & 66.4{\tiny$\pm$1.9} & 49.3{\tiny$\pm$1.7} & 36.1{\tiny$\pm$1.8} & 53.9{\tiny$\pm$1.3} & 62.4{\tiny$\pm$1.8} & 45.4{\tiny$\pm$1.7} & 31.9{\tiny$\pm$1.7} & 49.9{\tiny$\pm$1.4} & 9.7{\tiny$\pm$0.4} & 5.5{\tiny$\pm$0.3} & 5,012{\tiny$\pm$121} & 21.5{\tiny$\pm$1.9} \\
      AutoDev~\cite{tufano2024autodev} & 79.2{\tiny$\pm$1.8} & 61.0{\tiny$\pm$1.6} & 48.2{\tiny$\pm$1.7} & 66.2{\tiny$\pm$1.5} & 73.2{\tiny$\pm$1.7} & 55.1{\tiny$\pm$1.5} & 42.5{\tiny$\pm$1.6} & 60.3{\tiny$\pm$1.5} & 69.5{\tiny$\pm$1.6} & 51.4{\tiny$\pm$1.5} & 38.3{\tiny$\pm$1.8} & 56.4{\tiny$\pm$1.3} & 6.8{\tiny$\pm$0.3} & 3.9{\tiny$\pm$0.2} & 3,511{\tiny$\pm$96} & 19.2{\tiny$\pm$1.6} \\
      Claude~Code~\cite{anthropic_claudecode2026} & 86.0{\tiny$\pm$1.4} & 71.0{\tiny$\pm$1.4} & 53.0{\tiny$\pm$1.7} & 73.5{\tiny$\pm$1.3} & 78.0{\tiny$\pm$1.5} & 60.0{\tiny$\pm$1.2} & 43.0{\tiny$\pm$1.6} & 64.1{\tiny$\pm$1.4} & 75.4{\tiny$\pm$1.5} & 57.3{\tiny$\pm$1.4} & 40.3{\tiny$\pm$1.3} & 61.4{\tiny$\pm$1.1} & 12.4{\tiny$\pm$0.4} & 5.8{\tiny$\pm$0.3} & 5,210{\tiny$\pm$130} & 13.5{\tiny$\pm$1.3} \\
      Codex~\cite{openai_codex2026} & 83.0{\tiny$\pm$1.6} & 67.0{\tiny$\pm$1.5} & 49.0{\tiny$\pm$1.7} & 70.0{\tiny$\pm$1.4} & 74.0{\tiny$\pm$1.6} & 56.0{\tiny$\pm$1.6} & 39.0{\tiny$\pm$1.8} & 60.1{\tiny$\pm$1.3} & 71.4{\tiny$\pm$1.2} & 53.4{\tiny$\pm$1.4} & 36.5{\tiny$\pm$1.9} & 57.5{\tiny$\pm$1.5} & 11.2{\tiny$\pm$0.3} & 5.1{\tiny$\pm$0.2} & 4,820{\tiny$\pm$124} & 14.2{\tiny$\pm$1.4} \\
  \midrule
      \rowcolor{gray!10}
      \textbf{\sysmethod} & \textbf{97.3{\tiny$\pm$0.8}} & \textbf{78.5{\tiny$\pm$0.9}} & \textbf{61.2{\tiny$\pm$1.1}} & \textbf{82.9{\tiny$\pm$1.0}} & \textbf{94.0{\tiny$\pm$0.8}} &
\textbf{72.3{\tiny$\pm$0.9}} & \textbf{54.8{\tiny$\pm$1.0}} & \textbf{77.9{\tiny$\pm$0.9}} & \textbf{92.5{\tiny$\pm$0.9}} & \textbf{70.3{\tiny$\pm$0.7}} & \textbf{52.3{\tiny$\pm$1.2}} & \textbf{76.0{\tiny$\pm$0.7}} &  \textbf{8.7{\tiny$\pm$0.2}} & \textbf{4.9{\tiny$\pm$0.1}} & \textbf{4,481{\tiny$\pm$83}} & \textbf{7.7{\tiny$
\pm$0.8}} \\
  \bottomrule
  \end{tabular}
\end{threeparttable}}
\label{tab:rq1}
\vspace{-2.0ex}
\end{table*}

\subsection{Metrics}

As defined in $\S$~\ref{sec:instance}, each \sysbench instance is stored as a four-part record: a code pair $(C_{\text{bug}}, C_{\text{fix}})$, the controlled context $\mathcal{R}$, a pinned toolchain configuration $\boldsymbol{\theta}$, and the two-tier oracle protocol. For metric definitions, we only need the executable repair task induced by this record. We therefore view each instance under evaluation as
$x=\bigl(c^{(0)},\mathcal{R},\mathcal{O}_1,\mathcal{O}_2\bigr)$, where $c^{(0)} = C_{\text{bug}}$, $\mathcal{R}$ is the controlled context used for verification, and
$\mathcal{O}_1,\mathcal{O}_2$ are the Tier-1 and Tier-2 oracles instantiated under the toolchain
$\boldsymbol{\theta}$. Given a candidate $c$, the oracles return verifiable outcomes and diagnostics:
\begin{equation*}
\begin{aligned}
    \mathcal{O}_1(c;\mathcal{R}) \in \{\texttt{pass}, \texttt{fail}\}\times \mathcal{D}_1,\\
    \mathcal{O}_2(c;\mathcal{R}) \in \{\texttt{pass}, \texttt{fail}\}\times \mathcal{D}_2,
\end{aligned}
\end{equation*}
where $\mathcal{D}_1$ contains \texttt{opa check} reports and $\mathcal{D}_2$ contains failing-test logs from \texttt{opa test}. Tier-2 is evaluated only after Tier-1 passes.

Although Tier-2 provides an executable evaluation of policy behavior, the available tests may not cover all intended policy semantics. To strengthen this evaluation, we adopt the LLM-as-a-Judge paradigm~\cite{gu2026survey,he2026llm,jiang2026codejudgebench} and use GPT-5.4~\cite{openai_gpt54_2026} as a binary semantic judge. The judge is invoked only after a candidate passes Tier-2 and does not participate in repair generation. It classifies the repair as either \texttt{preserved} or \texttt{changed}, based on whether the repair fixes the target defect without introducing unintended changes to the policy behavior. The developer-written fix is provided only to the judge as a reference for the intended behavior.

The repair methods generate a candidate sequence $c^{(0)}, \dots, c^{(T)}$ 
within $T$ iterations. For an instance passing both Tier-1 and Tier-2, let \(\widehat{c}(x)\) denote the Tier-2-passing candidate returned by the repair method. We use \(\mathcal{J}^{\checkmark}(\widehat{c}(x);x,C_{\text{fix}})\) to denote that the judge classifies this candidate as preserved. Using the indicator \(\mathbb{I}\), we define:
\begin{equation*}
\begin{split}
\text{Pass@Tier-1}(x) &= \mathbb{I}\bigl[\exists\, t {\le} T{:}\ \mathcal{O}_1^{\checkmark}(c^{(t)})\bigr]\\
\text{Pass@Tier-2}(x) &= \mathbb{I}\bigl[\exists\, t {\le} T{:}\ \mathcal{O}_1^{\checkmark}(c^{(t)}) \land \mathcal{O}_2^{\checkmark}(c^{(t)})\bigr]\\
\text{JudgePass}(x)
& 
=\mathbb{I}\bigl[\text{Pass@Tier-2}(x)=1\land\mathcal{J}^{\checkmark}(\widehat{c}(x);x,C_{\text{fix}})\bigr]
\end{split}
\end{equation*}
Pass@Tier-1 measures syntax-level success, Pass@Tier-2 measures test-based semantic success, and JudgePass further requires the repaired policy to preserve the intended policy behavior. In our paper, repair success rates reported for Rego in RQ2-RQ4 and RQ6 refer to Pass@Tier-2. RQ1 additionally reports Pass@Tier-1 and JudgePass, while RQ5 uses the native oracle of each DSL.

To quantify trajectory stability, we track the number of newly introduced errors during the repair process. Let $V_{t}$ denote the total violation count at step $t$. We measure the regression magnitude at each step as the increase in errors compared to the previous step. The overall regression score is defined as the average number of new errors introduced per round:
\begin{align*}
\text{Regress}(t) &= max(0, V_{t}-V_{t-1}),\\
\text{Regr.\%} &= \frac{1}{T}\sum_{t=1}^{T}\text{Regress}(t)
\end{align*}

Under this definition, revealing a deeper-stage error after clearing an earlier blocker is treated as progress rather than regression.


\section{Experimental Results}
\subsection{RQ1: Overall Performance}

\begin{figure}
    \centering
    \scalebox{0.95}{
        \includegraphics[width=0.90\linewidth]{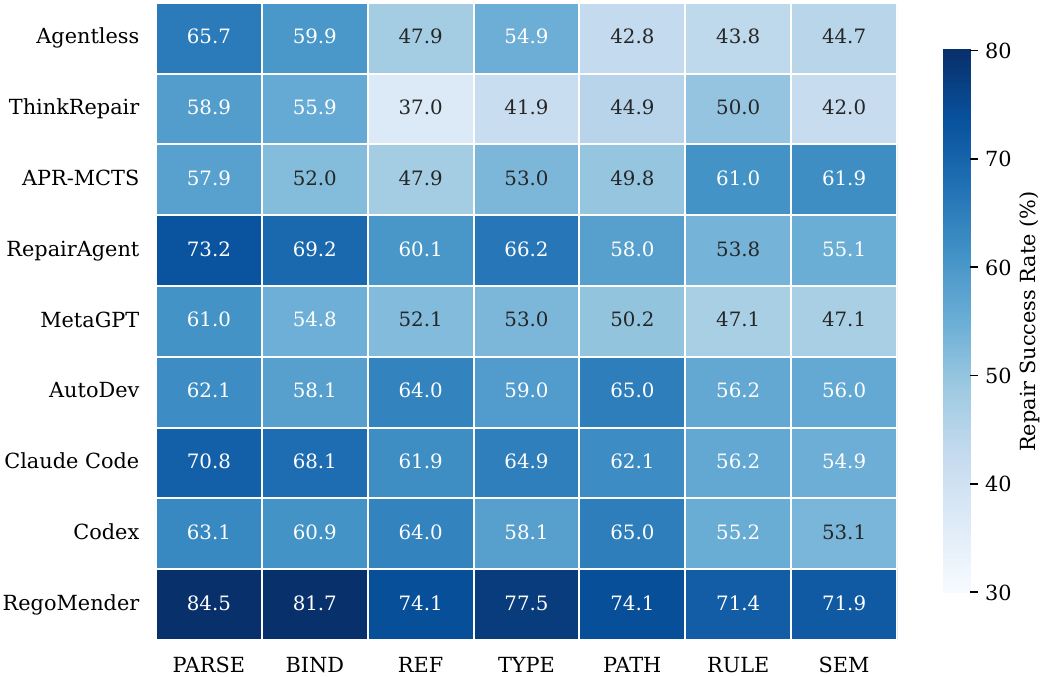}
    }
    \vspace{-2.5ex}
    \caption{Repair Success Rates for Different Defect Types.}
    \vspace{-4.0ex}
    \label{fig:heatmap}
\end{figure}

Table~\ref{tab:rq1} and Figure~\ref{fig:heatmap} present the comparative evaluation results on RegoBench across different difficulty levels and defect types, using DeepSeek-V3.1-Terminus~\cite{liu2024deepseek} as the backbone for all baselines.

Among the standalone baselines, RepairAgent performs the best, achieving 70.4\% on Pass@Tier-1 and 64.5\% on Pass@Tier-2, clearly outperforming APR-MCTS, Agentless, and ThinkRepair. This suggests that structured tool use is far more effective for Rego repair than prompting or search alone. However, \sysmethod surpasses RepairAgent by a substantial margin of 12.5\% on Pass@Tier-1 and 13.4\% on Pass@Tier-2. This gap shows that simply providing tools to a single agent is not enough; successful repair depends heavily on dynamic orchestration that can effectively coordinate context gathering, error diagnosis, and patch generation.

This reliance on proper orchestration becomes even clearer when comparing RegoMender with general-purpose agent systems. Multi-agent strategies such as AutoDev and MetaGPT achieve Pass@Tier-2 rates of 60.3\% and 53.9\%, respectively. The off-the-shelf coding agents perform better, with Claude Code reaching 64.1\% and Codex reaching 60.1\%. Nevertheless, RegoMender exceeds them by 13.8\% and 17.8\%, respectively. These results suggest that general-purpose systems have difficulty handling chained Rego errors without a domain-specific workflow. In contrast, RegoMender features a tailored orchestration mechanism that coordinates the Analyzer and Repairer according to repair progress, allowing the system to make suitable task decisions and resolve complex errors.

To assess the reliability of JudgePass, three senior developers independently label a stratified sample of 300 Tier-2-passing repairs. After resolving disagreements, GPT-5.4 agrees with the final expert labels on 96.7\% of the sampled repairs, with a Cohen’s $\kappa$~\cite{cohen1960coefficient} of 0.81. JudgePass remains close to Pass@Tier-2 across all evaluated methods, with differences ranging from 1.9\% to 4.3\%. For RegoMender, JudgePass reaches 76.0\%, only 1.9\% below its Pass@Tier-2 rate of 77.9\%. This corresponds to approximately 97.6\% of its Tier-2-passing repairs being judged to preserve the intended policy behavior. The strong agreement with expert labels supports the reliability of JudgePass, while the small gap between the two metrics supports the use of \texttt{opa test} as an executable proxy for Rego semantics. 

Figure~\ref{fig:heatmap} compares the Pass@Tier-2 rates of all baselines across the seven defect types. RegoMender ranks first for every defect type, outperforming the strongest baseline for each defect type by 9.1\% to 12.5\%. The largest improvement occurs for BIND, where RegoMender achieves 81.7\%, compared with 69.2\% for RepairAgent. It also improves PARSE and TYPE by 11.3\%. For REF and PATH, RegoMender reaches 74.1\%, exceeding the strongest baselines by 10.1\% and 9.1\%, respectively. Although RULE and SEM remain the most challenging, RegoMender achieves 71.4\% and 71.9\%, outperforming APR-MCTS by 10.4\% and 10\%. These results show that the overall improvement is not limited to parsing and binding errors but extends to dependency-related, structural, and semantic defects.

In terms of stability, \sysmethod achieves the lowest regression rate across all evaluated methods at just 7.7\%. Claude Code has the lowest regression rate among the baselines at 13.5\%, which is still 5.8\% higher than \sysmethod. This gap proves that our method rarely introduces more defects when modifying code. Regarding efficiency, \sysmethod naturally requires more budgets (\$8.7) and time (4.9h) than most baselines due to its multi-turn verification workflow. However, this added cost is a highly worthwhile trade-off. It directly translates into an absolute 13.4\% improvement in overall Tier-2 success over the strongest baseline, making the moderate increase in effort entirely justified.

\begin{table*}[!t]
  \centering
  \captionsetup{skip=4pt}
  \renewcommand{\arraystretch}{0.90}
  \caption{Performance Comparison of Repair Methods on RegoBench across Different Backbones.}
  \label{tab:backbone}
  \setlength{\tabcolsep}{0.70mm}
  \scalebox{0.80}{
  \begin{threeparttable}
    \newcommand{\mrcell}[1]{\textcolor{black}{#1}}
    \begin{tabular}{l|c|cccccc|cccccc|cccccc}
    \toprule
    \multirow{2}{*}{\textbf{Models}} & 
    \multirow{2}{*}{\textbf{Size}} & 
    \multicolumn{6}{c}{\textbf{ThinkRepair}} & 
    \multicolumn{6}{c}{\textbf{RepairAgent}} & 
    \multicolumn{6}{c}{\textbf{RegoMender}} \\
    \cmidrule(lr){3-8} \cmidrule(lr){9-14} \cmidrule(lr){15-20}
    
    & & \textit{Easy} & \textit{Medium} & \textit{Hard} & \textit{Overall} & \textit{Budget} & \textit{Time}
    & \textit{Easy} & \textit{Medium} & \textit{Hard} & \textit{Overall} & \textit{Budget} & \textit{Time} 
    & \textit{Easy} & \textit{Medium} & \textit{Hard} & \textit{Overall} & \textit{Budget} & \textit{Time} \\
    \midrule
    
    \multicolumn{20}{l}{\textit{\textbf{Closed-Source LLMs}}} \\
    GPT-4.1~\cite{openai_gpt41_2025} & - & 72.3 & 47.1 & 28.4 & 54.0 & 21.0 & 3.2 & 58.7 & 40.2 & 32.1 & 46.6 & 26.4 & 4.1 & 89.2 & 72.4 & 58.3 & 76.6 & 47.3 & 7.3 \\
    Claude-3.5-Sonnet~\cite{anthropic_claude35_sonnet_2024} & - & 82.1 & 55.3 & 33.2 & 62.1 & 27.0 & 2.4 & 72.4 & 58.1 & 48.3 & 62.2 & 33.8 & 3.0 & 92.3 & 76.2 & 60.1 & 79.7 & 60.8 & 5.4 \\
    Gemini-2.5-Flash~\cite{comanici2025gemini} & - & 35.2 & 22.1 & 12.3 & 25.7 & 10.0 & 2.7 & 80.3 & 62.1 & 52.2 & 67.9 & 12.6 & 3.4 & 90.1 & 70.3 & 57.2 & 76.1 & 22.5 & 6.1 \\
    \mrcell{GPT-5.4~\cite{openai_gpt54_2026}} & \mrcell{--} & \mrcell{81.5} & \mrcell{58.4} & \mrcell{\underline{38.6}} & \mrcell{64.2} & \mrcell{17.8} & \mrcell{3.0} & \mrcell{\underline{84.6}} & \mrcell{65.8} & \mrcell{53.4} & \mrcell{71.5} & \mrcell{22.5} & \mrcell{3.6} & \mrcell{95.2} & \mrcell{76.5} & \mrcell{61.0} & \mrcell{81.1} & \mrcell{41.2} & \mrcell{6.4} \\
    \mrcell{Claude-4.5-Sonnet~\cite{anthropic_claude45_sonnet_2025}} & \mrcell{--} & \mrcell{\underline{86.6}} & \mrcell{\underline{60.7}} & \mrcell{38.4} & \mrcell{\underline{67.1}} & \mrcell{28.1} & \mrcell{2.5} & \mrcell{83.6} & \mrcell{\underline{67.1}} & \mrcell{\underline{55.7}} & \mrcell{\underline{71.8}} & \mrcell{35.1} & \mrcell{3.1} & \mrcell{\underline{96.8}} & \mrcell{\underline{77.8}} & \mrcell{\underline{61.2}} & \mrcell{\underline{82.4}} & \mrcell{62.4} & \mrcell{5.6} \\
    \mrcell{Gemini-3-Flash~\cite{google_deepmind_gemini3_flash_2025}} & \mrcell{--} & \mrcell{37.0} & \mrcell{23.5} & \mrcell{13.5} & \mrcell{27.2} & \mrcell{10.9} & \mrcell{2.9} & \mrcell{83.5} & \mrcell{64.6} & \mrcell{54.3} & \mrcell{70.6} & \mrcell{13.6} & \mrcell{3.6} & \mrcell{93.3} & \mrcell{72.2} & \mrcell{58.5} & \mrcell{78.4} & \mrcell{23.1} & \mrcell{6.2} \\
    \midrule
    
    \multicolumn{20}{l}{\textit{\textbf{Open-Source LLMs}}} \\
    Qwen2.5-32B-Instruct~\cite{qwen2025qwen25technicalreport} & 32B & 21.3 & 14.1 & 7.2 & 15.7 & \underline{2.0} & 1.8 & 38.4 & 18.2 & 14.1 & 26.3 & \underline{2.5} & 2.3 & 84.2 & 64.3 & 50.1 & 69.9 & \underline{4.5} & 4.1 \\
    Llama-3.3-70B~\cite{llama33_modelcard} & 70B & 15.2 & 10.3 & 5.1 & 11.3 & 3.0 & \underline{1.4} & 48.3 & 28.1 & 22.2 & 35.8 & 3.8 & \underline{1.7} & 86.1 & 66.2 & 50.3 & 71.4 & 6.8 & \underline{3.1} \\
    DeepSeek-V3.1-Terminus~\cite{liu2024deepseek} & 685B & 65.1 & 42.0 & 24.1 & 48.2 & 3.9 & 2.2 & 78.5 & 58.4 & 45.8 & 64.5 & 5.2 & 2.9 & 94.0 & 72.3 & 54.8 & 77.9 & 8.7 & 4.9 \\
    \mrcell{DeepSeek-V4-Pro~\cite{deepseek_v4_2026}} & \mrcell{1.6T} & \mrcell{70.4} & \mrcell{46.5} & \mrcell{27.5} & \mrcell{52.5} & \mrcell{13.6} & \mrcell{1.9} & \mrcell{\underline{84.6}} & \mrcell{63.3} & \mrcell{50.0} & \mrcell{69.7} & \mrcell{17.4} & \mrcell{2.4} & \mrcell{95.3} & \mrcell{74.0} & \mrcell{56.0} & \mrcell{79.5} & \mrcell{28.2} & \mrcell{4.1} \\
    
    \bottomrule
    \end{tabular}
    \begin{tablenotes}
        \scriptsize
        \item \textit{Note:} Repair success rates (Easy/Medium/Hard/Overall) are in \%. Budget is in USD (\$). Time is in hours (h). Underlines mark the best values within each model group.
    \end{tablenotes}
	  \end{threeparttable}}
  \vspace{-3.0ex}
\end{table*}

\begin{rqanswerbox}
\textbf{Answer to RQ1:} \sysmethod outperforms all baselines, achieving a 13.4\% higher Pass@Tier-2 rate than the best baseline. 
Its JudgePass rate reaches 76.0\%, only 1.9\% below Pass@Tier-2, supporting \texttt{opa test} as a reliable proxy for semantics. \sysmethod also ranks first across seven defect types and achieves the lowest regression rate of 7.7\%, demonstrating broad repair effectiveness and stability across Rego defects.
\end{rqanswerbox}

\subsection{RQ2: Impact of LLM Backbones}

Table~\ref{tab:backbone} evaluates \sysmethod across ten backbone models, including six closed-source models (GPT-4.1~\cite{openai_gpt41_2025}, Claude-3.5-Sonnet~\cite{anthropic_claude35_sonnet_2024}, Gemini-2.5-Flash~\cite{comanici2025gemini}, GPT-5.4~\cite{openai_gpt54_2026}, Claude-4.5-Sonnet~\cite{anthropic_claude45_sonnet_2025}, Gemini-3-Flash~\cite{google_deepmind_gemini3_flash_2025}) and four open-source models (Qwen2.5-32B~\cite{qwen2025qwen25technicalreport}, Llama-3.3-70B~\cite{llama33_modelcard}, DeepSeek-V3.1-Terminus~\cite{liu2024deepseek}, DeepSeek-V4-Pro~\cite{deepseek_v4_2026}). The results demonstrate that our framework improves repair success across model sizes and types.

Our framework significantly empowers smaller open-source models. Under the ThinkRepair baseline, smaller models struggle with complex logic. For instance, Qwen2.5-32B and Llama-3.3-70B achieve success rates of only 7.2\% and 5.1\% in the Hard category. This indicates that without guidance, they lack the reasoning depth for difficult semantic errors. However, when equipped with our orchestrated workflow, their performance improves dramatically. Qwen2.5-32B sees its success rate in the Hard category jump to 50.1\%. This result is competitive with the much larger Gemini-2.5-Flash (57.2\%) and approaches GPT-4.1 (58.3\%). These results suggest that our framework helps smaller models solve difficult tasks by decomposing them into manageable steps. The benefits also extend to stronger models. Claude-4.5-Sonnet achieves an overall success rate of 82.4\%, improving upon the strongest baseline result of 71.8\% by 10.6 percentage points, while Gemini-3-Flash reaches 78.4\%, compared with 70.6\% for the strongest baseline.

In terms of cost efficiency, DeepSeek-V3.1-Terminus provides the most balanced solution. Although Claude-4.5-Sonnet achieves the highest hard-instance success rate of 61.2\%, it requires a budget of \$62.4. In contrast, DeepSeek-V3.1 achieves a comparable success rate of 54.8\% but costs only \$8.7. This represents an approximately 86\% cost reduction while keeping the repair success rate close to the best result. DeepSeek-V4-Pro further improves the overall success rate to 79.5\%, but its cost of \$28.2 is more than three times that of DeepSeek-V3.1. These results show that RegoMender can remain effective under strict budget limits.

\begin{rqanswerbox}
\textbf{Answer to RQ2:} RegoMender demonstrates robust generalizability across all tested models, effectively bridging the performance gap between open-source and closed-source LLMs. Claude-4.5-Sonnet achieves the highest repair success rate, while DeepSeek-V3.1-Terminus provides the best balance between repair success and cost, reaching an overall success rate of 77.9\% at a cost of only \$8.7.
\end{rqanswerbox}

\begin{figure}[t]
  \centering
  \captionsetup{skip=4pt}
  \scalebox{0.90}{
    \includegraphics[width=0.90\linewidth]{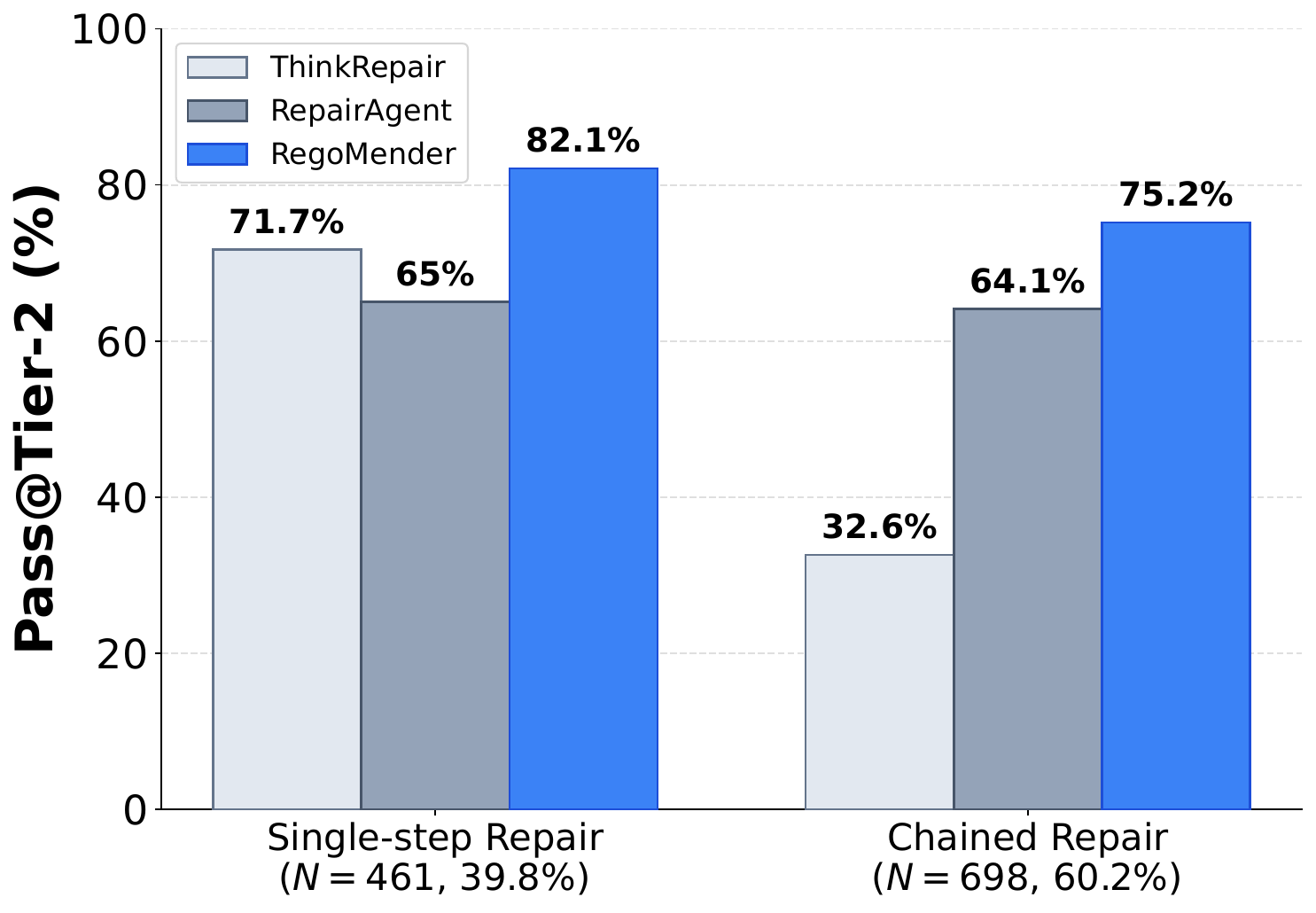}
  }
  \vspace{-1.5ex}
  \caption{Repair Success: Single-step vs. Chained.}
  \vspace{-4.0ex}
  \label{fig:chain-repair}
\end{figure}

\begin{figure*}[tbp]
  \centering
  \captionsetup{skip=4pt}
  \scalebox{0.95}{
    \includegraphics[width=0.90\textwidth]{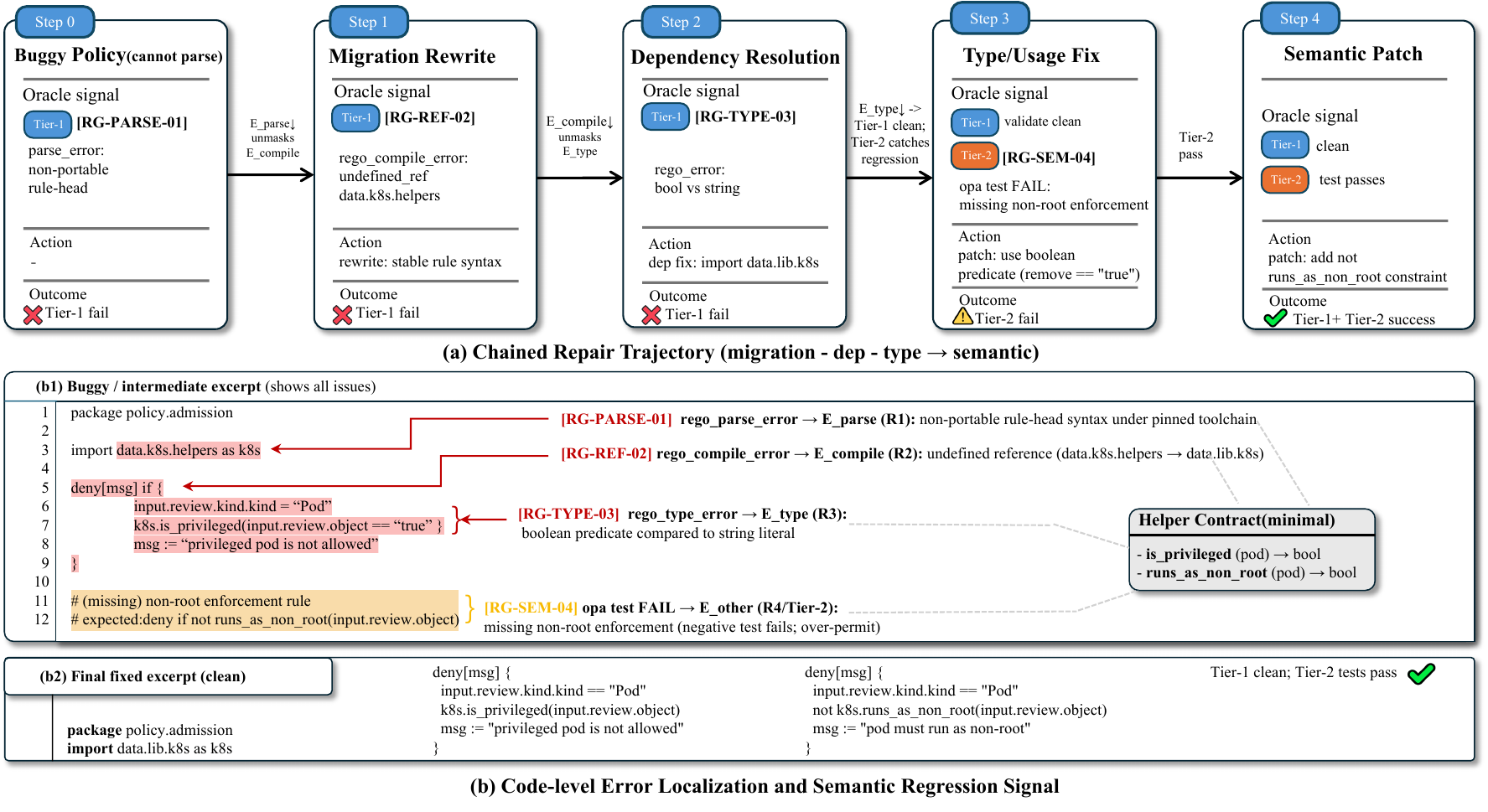} 
  }
  \vspace{-2.0ex}
  \caption{Chained Repair Process. \sysmethod resolves syntax errors then passes semantic tests to produce a verified patch.}
  \label{fig:case}
  \vspace{-3.0ex}
\end{figure*}

\vspace{-2.0ex}
\subsection{RQ3: Efficacy of Multi-agent Orchestration}

To better understand the practical challenges of Rego repair, we analyze single-step and chained repairs separately. For each category, we compare the ThinkRepair and RepairAgent baselines against \sysmethod using the overall success rate.

Figure~\ref{fig:chain-repair} shows that \sysmethod achieves the best performance in both categories. On single-step repairs, ThinkRepair, RepairAgent, and \sysmethod achieve success rates of 71.7\%, 65.0\%, and 82.1\%, respectively. On chained repairs, their success rates are 32.6\%, 64.1\%, and 75.2\%. ThinkRepair shows the largest drop on chained repairs, while RepairAgent remains relatively stable. Overall, \sysmethod consistently performs best in both settings, indicating stronger adaptability to single-step and chained Rego repair tasks.

Figure~\ref{fig:case} shows a complete chained repair case where hidden failures progressively emerge as verification advances. The buggy code first fails syntax parsing due to an unsupported rule syntax (\texttt{RG-PARSE-01}), prompting an initial syntax migration to normalize the format. This migration then exposes a compilation blocker (\texttt{RG-REF-02}). The policy imports an outdated helper namespace (\texttt{data.k8s.helpers}) that has moved to \texttt{data.lib.k8s}, causing an undefined reference. After resolving the dependency by importing the correct module, the verifier reveals a type usage issue (\texttt{RG-TYPE-03}) where a boolean helper is incorrectly compared against a string literal (e.g., \texttt{== "true"}). This error requires a rewrite into a boolean predicate. The resulting code passes all syntax checks, but semantic tests still fail with a security regression (\texttt{RG-SEM-04}). Negative tests indicate the policy is missing a required non root enforcement rule, which would allow insecure pods. The repair completes only after adding the missing semantic constraint, namely a \texttt{deny} rule that enforces \texttt{not k8s.runs\_as\_non\_root(...)}.

This scenario demonstrates why chained repair is fundamentally different from single-step editing. Early parsing and compilation failures can easily hide downstream type and security issues, and even a structurally sound policy can still fail semantic security tests. Accordingly, our framework strictly follows verification signals in the correct logical sequence. It progresses from syntax migration and dependency resolution, to type correction, and finally to a semantic patch. The framework continues iterating until the policy successfully passes all verification stages.

\begin{rqanswerbox}
\textbf{Answer to RQ3:} \sysmethod achieves the highest success rate on both single-step and chained repairs. The results show that different methods vary substantially in their robustness to chained repair scenarios: ThinkRepair drops sharply on chained repairs, whereas RepairAgent remains relatively stable. In comparison, \sysmethod consistently performs best in both settings, indicating stronger adaptability to single-step and chained Rego repair tasks.
\end{rqanswerbox}

\subsection{RQ4: Ablation Study}

\begin{table}[t]
    \centering
    \captionsetup{skip=4pt}
    \caption{Ablation Study on Tool Effectiveness.}
    \label{tab:tool_ablation}
    \renewcommand{\arraystretch}{0.90}
    \setlength{\tabcolsep}{1.5mm}
    \scalebox{0.95}{
        \begin{tabular}{lcc}
        \toprule
        \textbf{Tool} & \textbf{RepairAgent (\%)} & \textbf{\sysmethod (\%)} \\
        \midrule
        Full Toolkit & 64.5 & 77.9 \\
        \midrule
        w/o \textit{Context Scan} & \hspace{0.8ex}49.2$_{\textcolor{red}{(\downarrow 15.3)}}$ & \hspace{0.8ex}61.2$_{\color{red}{(\downarrow 16.7)}}$ \\
        w/o \textit{Patch} & \hspace{0.8ex}51.7$_{\textcolor{red}{(\downarrow 12.8)}}$ & \hspace{0.8ex}64.1$_{\textcolor{red}{(\downarrow 13.8)}}$ \\
        w/o \textit{Rollback} & --- & \hspace{0.8ex}66.4$_{\textcolor{red}{(\downarrow 11.5)}}$ \\
        w/o \textit{Rewrite} & 56.4$_{\textcolor{red}{(\downarrow 8.1)}}$ & 68.6$_{\textcolor{red}{(\downarrow 9.3)}}$ \\
        w/o \textit{Error Triage} & 55.1$_{\textcolor{red}{(\downarrow 9.4)}}$ & \hspace{0.8ex}67.2$_{\textcolor{red}{(\downarrow 10.7)}}$ \\
        \bottomrule
        \end{tabular}
    }
\vspace{-3.0ex}
\end{table}

Based on the ablation study in Table~\ref{tab:tool_ablation}, we analyze the contribution of each tool to the overall repair success. The results identify \texttt{Context Scan} as the most critical tool. Its removal causes a severe decline in the success rate, dropping by 15.3\% in the RepairAgent and 16.7\% in the \sysmethod. This decrease confirms that the primary challenge in Rego repair is resolving dependencies and import structures rather than merely fixing isolated syntax errors. Without scanning the broader context, the agent lacks the necessary information to understand the policy scope. Next in importance is the \texttt{Patch} tool, whose removal leads to a significant performance reduction of approximately 13\%. This indicates that applying precise edits is more effective than rewriting entire code blocks, which can inadvertently introduce new defects. Without Rollback, RegoMender drops from 77.9\% to 66.4\%, a decrease of 11.5\%. This result shows that Rollback helps the framework return to an earlier stable state when a repair attempt stalls or introduces new errors, preventing unsuccessful changes from affecting later repair steps. Furthermore, the exclusion of \texttt{Error Triage} and \texttt{Rewrite} results in performance losses of around 10\%. This demonstrates the necessity of filtering raw compiler messages and providing a fallback for complex logic changes. Overall, the consistent performance drops across all ablation configurations validate that every tool in our framework is essential for achieving high-performance repair.

\begin{table}[t]
\centering
\captionsetup{skip=4pt}
\caption{Cross-DSL Generalization Performance.}
\label{tab:cross_dsl}
\renewcommand{\arraystretch}{1.0}
\scalebox{0.85}{
    \begin{tabular}{lcccc}
    \toprule
    \textbf{DSL} & \textbf{N} & \textbf{Avg.Errors} & \textbf{RepairAgent (\%)} & \textbf{RegoMender (\%)} \\ 
    \midrule
    \textbf{Rego} & 1,159 & 6.2 & 64.5 & 77.9 \\
    HCL & 300 & 4.6 & 78.2 & 82.3 \\
    SQL & 100 & 1.0 & 84.0 & 84.0 \\ 
    \bottomrule
    \end{tabular}
}
\vspace{-4.0ex}
\end{table}

\begin{rqanswerbox}
\textbf{Answer to RQ4:} 
The ablation study demonstrates that \texttt{Context Scan} and \texttt{Patch} are the two most critical components. Removing either of these tools leads to the most significant performance drops. This finding confirms that effective Rego repair relies on resolving dependencies and making precise adjustments based on the broader scope.
\end{rqanswerbox}

\subsection{RQ5: Generalizability to Other Languages}

To investigate whether our framework generalizes beyond Rego, we expand our evaluation to include two other domain-specific languages, HCL and SQL, using their respective native oracles like \texttt{tf fmt} and standard SQL parsers. Applying the same filtering steps used for Rego, we collect 300 real-world HCL defects from 23 popular GitHub repositories and 100 SQL defects from 50 repositories.

Compared to Rego, both HCL and SQL offer clearer initial error messages, but they differ greatly in repair complexity. HCL is mainly used for cloud infrastructure. While its native tools easily point out the exact line of a syntax issue, a large portion of HCL defects cannot be repaired with a single change. Because infrastructure code is highly connected, fixing one resource often exposes an error in another, requiring multi-step repairs. In contrast, SQL queries are typically self-contained. Their defects are mostly simple syntax mistakes, like a missing keyword, which usually require only one local edit without cross-file dependencies. HCL and SQL provide a perfect testbed: HCL tests multi-step repairs with clear feedback, while SQL tests simple, single-step repairs.


Table~\ref{tab:cross_dsl} indicates that the benefit of orchestration depends on the repair profile induced by the language and its oracle feedback, rather than remaining constant across DSLs. On Rego, as defects typically require repository-specific reasoning (e.g., cross-file dependencies), \sysmethod improved over RepairAgent (77.9\% vs.\ 64.5\%), consistent with our main findings. On HCL, \sysmethod still outperforms RepairAgent (82.3\% vs.\ 78.2\%), suggesting that orchestration remains helpful when oracle feedback is relatively precise but the repair space is still more complex than SQL. On SQL, as defects are mostly lightweight parse failures and the oracle is simpler, both methods performed identically (84.0\%).

\begin{rqanswerbox}
\textbf{Answer to RQ5:} Overall, our results suggest that generalizability depends on DSL characteristics. Orchestration methods are most effective when repair tasks rely on multi-step reasoning and feedback localization is ambiguous.
\end{rqanswerbox}

\subsection{RQ6: Real-world Vulnerability Repair}

To assess reliability in security critical scenarios, we evaluate the ThinkRepair, the RepairAgent, and \sysmethod on 132 real-world vulnerabilities. We construct this specific subset by filtering RegoBench for instances linked to known CVE records.

\begin{table}[t]
    \centering
    \captionsetup{skip=4pt}
    \caption{Security Policies Repair Rates by CWE Category.}
    \label{tab:cwe_repair_rates}
    \renewcommand{\arraystretch}{1.15}
    \scalebox{0.72}{
        \begin{tabular}{llcccc}
        \toprule
        \textbf{CWE} & \textbf{Description} & \textbf{N} & \textbf{ThinkRepair} & \textbf{RepairAgent} & \textbf{\sysmethod} \\
        \midrule
        CWE-284 & Access Control & 31 & 28.1\% & 71.9\% & 87.5\% \\
        CWE-250 & Excess Privilege & 28 & 27.6\% & 72.4\% & 89.7\% \\
        CWE-269 & Privilege Mgmt & 25 & 26.9\% & 69.2\% & 88.5\% \\
        CWE-732 & Permission Assign  & 21 & 27.3\% & 68.2\% & 82.0\% \\
        CWE-863 & Incorrect Authz & 19 & 20.0\% & 60.0\% & 80.0\% \\
        Other   & Mixed Types &  9 & 30.0\% & 50.0\% & 90.0\% \\
        \midrule
        \textbf{Total} & --- & \textbf{132} & \textbf{26.6\%} & \textbf{67.6\%} & \textbf{86.4\%} \\
        \bottomrule
        \end{tabular}
    }
\vspace{-5.0ex}
\end{table}

Table~\ref{tab:cwe_repair_rates} presents repair performance across different Common Weakness Enumerations (CWEs). The results highlight that direct prompting is insufficient for security patches, solving only 26.6\%. However, agentic workflows significantly improve these results. Compared to the 67.6\% success rate of the RepairAgent, our method reaches 86.4\%, delivering a clear 18.8\% overall gain. Breaking down this overall improvement by vulnerability type reveals two key trends. First, \sysmethod shows consistent improvements across major categories. It outperforms the RepairAgent by 15.6\% in Access Control and 19.3\% in Privilege Management. This advantage expands to 20.0\% for Incorrect Authorization due to complex logic. Second, our method proves robust on diverse cases. It achieves a 40.0\% performance boost in the Other category. This result confirms that a multi-agent system handles diverse vulnerability types much more effectively than a single agent.

\begin{rqanswerbox}
\textbf{Answer to RQ6:} \sysmethod significantly outperforms baselines in repairing real-world vulnerabilities, achieving 86.4\% success rate. More importantly, it is more reliable than RepairAgent when dealing with diverse and mixed vulnerability types.
\end{rqanswerbox}

\vspace{-1.0ex}
\section{Discussion}

\subsection{Failure Analysis}

To better understand the limitations of our approach, we present a qualitative analysis of failure instances where all evaluated methods failed to produce a valid repair.

\textbf{Unreachable Context.} 
Failures often occur when project dependencies cannot be fully resolved. In complex projects, a file might rely on private modules, external code, or dynamically generated files. Since static analysis tools cannot access these missing parts, they provide an incomplete context. To fill these gaps, the model generates references or functions that look correct but are invalid. This keeps the system stuck in a loop of failed repair attempts, showing a limitation of relying solely on static analysis.

\textbf{Iteration Limitation.} 
Beyond incomplete context, strict limits on repair iterations present another major hurdle. When a task involves multiple linked errors, fixing one bug often exposes another, creating a chain of issues that requires many attempts to fully resolve. Given the need to control costs, we set a maximum iteration limit, which can cause the \sysmethod repair process to end just before the code is verified. Consequently, successfully repairing deeply interconnected errors under a strict iteration budget remains a major challenge.

\textbf{Semantic Ambiguity.}
Finally, failures can arise from coding conventions rather than simple syntax errors. A policy can be implemented in multiple valid ways, but only specific patterns align with the repository's established style, such as the choice between \texttt{deny} and \texttt{allow} layouts. Without clear examples, the model may generate a patch that is technically correct but fails strict stylistic tests or is rejected during human review.

\vspace{-2.0ex}
\subsection{Threats to Validity}

\noindent\textbf{Internal Validity.} One primary threat is that a repair may satisfy \texttt{opa check} and \texttt{opa test} while failing to fully preserve the original policy intent, particularly when test coverage is incomplete and misses edge cases.
Another threat involves potential noise or data contamination in mined repair data, as commit histories may include refactoring or deletions unrelated to bug fixing. We mitigate these threats in two ways. First, we construct \sysbench using a rigorous three-stage filtering pipeline and a fixed OPA toolchain to ensure each instance is reproducible and evaluated against native verification signals. Second, to strengthen semantic evaluation beyond the available tests, we combine two-tier verification and regression monitoring with JudgePass, which examines Tier-2-passing repairs for unintended changes to policy behavior. We further assess the reliability of JudgePass against expert labels on a stratified sample. The close agreement between Pass@Tier-2 and JudgePass, together with the lower regression rate of \sysmethod, provides complementary evidence of semantic preservation and repair stability. Nevertheless, incomplete tests and specifications may still leave some semantic differences undetected.

\noindent\textbf{External Validity.} A main external threat is that our study is centered on Rego, so the findings may not transfer equally well to other Policy-as-Code settings or to simpler DSLs with different oracle behavior. Another practical deployment threat is that real environments may involve different model choices, budget constraints, and security-sensitive scenarios. We partially address these concerns through broader evaluation. Across different LLM backbones, RegoMender consistently improves repair performance, including on smaller open-source models. Beyond Rego, the cross-DSL results on HCL and SQL suggest that the value of orchestration is not limited to one language, though the gain is larger when the repair requires multi-step reasoning. Finally, on 132 CVE-linked real-world vulnerability cases, RegoMender achieves strong results and clearly outperforms baselines. These findings support the practical relevance of our method, although further validation on more languages and production settings is still needed.

\vspace{-2.0ex}
\section{Related Work}
\label{related}

\vspace{-0.5ex}
\subsection{Automated Program Repair}
\vspace{-0.3ex}

Automated program repair has emerged as a promising technique to alleviate the manual debugging burden by automatically generating patches for software defects. Traditional APR methods typically rely on test suites as specifications and can be broadly categorized into search-based and constraint-based approaches. Search-based methods, such as GenProg~\cite{le2011genprog} and SimFix~\cite{jiang2018shaping}, formulate patch generation as a search problem over a syntactic mutation space. Conversely, constraint-based approaches, like SemFix~\cite{nguyen2013semfix} and Angelix~\cite{mechtaev2016angelix}, leverage symbolic execution to extract program semantics and synthesize patches via constraint solving. While these traditional methods have demonstrated empirical success on general-purpose programming languages like Java and C, their efficacy is often bottlenecked by the exponential search space and the scalability limitations of underlying solvers.

In recent years, the rapid advancement of deep learning has catalyzed a paradigm shift towards learning-based APR. Early data-driven approaches~\cite{gupta2017deepfix,tufano2019learning,lutellier2020coconut,chen2019sequencer} treated patch generation as a Neural Machine Translation (NMT) task, learning bug-fixing patterns from large-scale historical commits. More recently, large language models pre-trained on massive code corpora have exhibited remarkable capabilities in understanding code context and generating plausible patches through methods like few-shot prompting~\cite{xia2024automated,yin2024thinkrepair,kang2025explainable} or fine-tuning~\cite{li2024exploring}. These learning-based models have significantly improved the naturalness and correctness of generated patches, pushing the boundaries of automated debugging in modern software development~\cite{fan2023large}.

Despite these improvements, existing APR methods primarily focus on general-purpose languages like Java or Python, leaving a significant gap in the realm of domain-specific languages and declarative policies. Recent studies have explored related tasks in infrastructure repair and policy generation. InfraFix~\cite{saavedra2025infrafix} repairs IaC scripts using verification rules designed for infrastructure configurations rather than Rego policies. ARPaCCino~\cite{romeo2025arpaccino} generates Rego policies from natural-language requirements to address Terraform violations, while Prose2Policy~\cite{gupta2026prose2policy} translates natural-language access-control policies into executable Rego. Overall, these systems cover IaC repair and Rego synthesis, but do not address repository-level repair of existing faulty Rego policies using the OPA toolchain. Therefore, advancing APR in this domain requires novel frameworks designed specifically to capture the underlying logic and structural constraints of declarative policies.

\vspace{-2.0ex}
\subsection{Multi-agent System}
\vspace{-0.3ex}

Multi-agent Systems (MAS) have shown strong potential in software engineering~\cite{hu2025compileagent,bouzenia2025repairagent,huang2023agentcoder,islam2024mapcoder,becattini2025sallma,peng2025pwngpt} by decomposing complex task into collaborative workflows~\cite{he2024mas4se_survey}. Existing systems generally follow two main design paradigms. The first is role-based collaboration, where agents are assigned specialized responsibilities resembling human development teams, such as planning, coding, and reviewing, as exemplified by systems like ChatDev~\cite{qian2024chatdev} and MetaGPT~\cite{hong2023metagpt}, which aim to improve modularity and coordination in large-scale development workflows. The second paradigm focuses on tool-integrated execution, where agents operate in sandboxed environments and interact directly with external tools and verification infrastructure to support end-to-end task execution~\cite{tufano2024autodev,yang2024swe,zhang2024autocoderover,wang2024openhands}.

In the context of program repair, MAS naturally enables a division of labor by separating key stages such as fault localization, patch generation, and regression testing. It is often combined with search strategies to handle noisy or ambiguous feedback. However, existing MAS-based repair approaches still face significant challenges, particularly in terms of agent coordination and integration with external verifiers~\cite{cemri2025why_mas_fail,kong2025aegisautomatederrorgeneration,xia2024agentless}. These challenges are even more pronounced in verification-intensive domains such as Policy-as-Code. In such cases, using MAS to separate diagnosis from oracle execution proves especially beneficial, as it enables a more auditable and reproducible repair loop, while also mitigating the brittleness commonly associated with Rego policy repair~\cite{opa_policy_language,romeo2025arpaccino}.

\vspace{-2.0ex}
\section{Conclusion and Future Work}
\vspace{-0.3ex}

In this paper, we address the challenge of applying LLM-based repair to the strict reliability requirements of cloud-native Policy-as-Code. We focus on automated Rego repair, where success requires passing a rigorous verification toolchain rather than merely generating plausible code. Our experiments show that \sysmethod consistently outperforms standalone baselines and general-purpose agent systems, especially on complex, multi-step defects and security-critical issues, while reducing the risk of introducing new bugs. These results suggest that, for domain-specific language repair, system architecture can be more important than model capability alone, especially when it effectively uses repository context and verifier feedback. \sysbench and \sysmethod provide a foundation for building reliable autonomous DevSecOps agents.


Building on these findings, several directions merit exploration. First, expanding \sysbench to other Policy-as-Code languages would help assess the generality of our approach. Second, because production policies often depend on external modules and dynamic cluster states, combining static analysis with sandboxed execution could capture these dependencies and improve repair accuracy. Third, syntactically valid patches may still weaken security; automated negative tests and intent-preservation checks could prevent such repairs. Finally, future work should assess the cost–benefit trade-off of multi-agent collaboration for practical deployment.

\vspace{-2.0ex}
\section*{Data Availability Statement}
\vspace{-0.3ex}

To support reproducibility, we release a replication package containing: (i) the available portion of RegoBench, including defect instances and metadata, (ii) evaluation scripts and configuration files for reproducing the main experimental results, and (iii) sample execution traces illustrating the repair process. Artifacts are available anonymously at \url{https://doi.org/10.5281/zenodo.19230997}.

\begin{acks}

This research was partially supported by the National Cyber Security-National Science and Technology Major Project of China under Grant No. 2025ZD1500800.
\end{acks}

\bibliographystyle{ACM-Reference-Format}
\bibliography{main}

\end{document}